\pdfoutput=1
\documentclass[11pt]{article}

\usepackage{acl}

\usepackage{times}
\usepackage{latexsym}
\usepackage[T1]{fontenc}
\usepackage[utf8]{inputenc}
\usepackage{microtype}
\usepackage{inconsolata}
\usepackage{graphicx}
\usepackage{amsmath}
\usepackage{amsthm}
\usepackage{amssymb}
\usepackage{bbm}
\usepackage{booktabs}
\usepackage{array}
\usepackage{algorithm}
\usepackage{algorithmic}
\usepackage{multirow}
\usepackage{xcolor}
\usepackage{pgfplots}
\pgfplotsset{compat=1.18}
\usepgfplotslibrary{groupplots}
\usetikzlibrary{patterns,calc}
\definecolor{serA}{HTML}{2A78D6}
\definecolor{serB}{HTML}{EB6834}
\definecolor{inkmuted}{HTML}{52514E}
\definecolor{gridln}{HTML}{D8D8D4}
\definecolor{ctxdot}{HTML}{9A9A94}
\usepackage{soul}
\usepackage{url}

\newcommand{\spm}[1]{\,{\scriptsize$\pm$#1}}      
\newcommand{\dr}[1]{\,{\scriptsize(#1)}}  

\newcommand{\narrowtable}{\footnotesize\setlength{\tabcolsep}{2pt}}

\title{Adversarial Robustness of Retrieval-Based Fact-Checking: Detecting and Mitigating Evidence Contamination}

\author{
  Alamgir Munir Qazi$^{1}$ \quad
  John P. McCrae$^{2}$ \quad
  Jamal Abdul Nasir$^{1}$ \\
  $^{1}$School of Computer Science, University of Galway, Ireland \\
  $^{2}$Research Ireland Insight Centre and ADAPT Centre, University of Galway, Ireland \\
  \texttt{\{a.qazi1,jamal.nasir\}@universityofgalway.ie, john@mccr.ae}
}

\begin{document}

\maketitle
\begin{abstract} 
Retrieval-based fact-checking systems ground verification in large document collections, creating a critical vulnerability: adversarial evidence contamination. We demonstrate that factually inconsistent but contextually coherent evidence generated to satisfy both retrieval-similarity and linguistic-fluency constraints can systematically mislead verification models. Contaminating just 3.1--7.4\% of training corpora with five label-specific strategies causes adversarial content to penetrate up to 35.2\% of top-k retrievals, degrading verification accuracy by up to 46.9\% relative F1. Existing retrieval-stage defenses, validated against gradient-optimized or syntactically anomalous attacks, transfer poorly to this threat model: detection rates collapse and F1 recovery exhibits high variance across seeds. 

We propose GLAD (Geometric-Lexical Adversarial Defense), which fuses dense embedding peer-coherence with lexical distributional isolation. Across two benchmarks, two embedding models, and five random seeds, GLAD achieves the highest detection rates (70--86\%) and the most consistent F1 recovery, reclaiming 73--80\% of contamination-induced degradation on LIAR-RAW while approaching clean baselines on RAWFC. Crucially, unlike defenses that trade clean performance for robustness, GLAD imposes negligible cost on uncontaminated corpora.

\end{abstract}

\begin{figure*}[t]
  \centering
  \includegraphics[width=\textwidth]{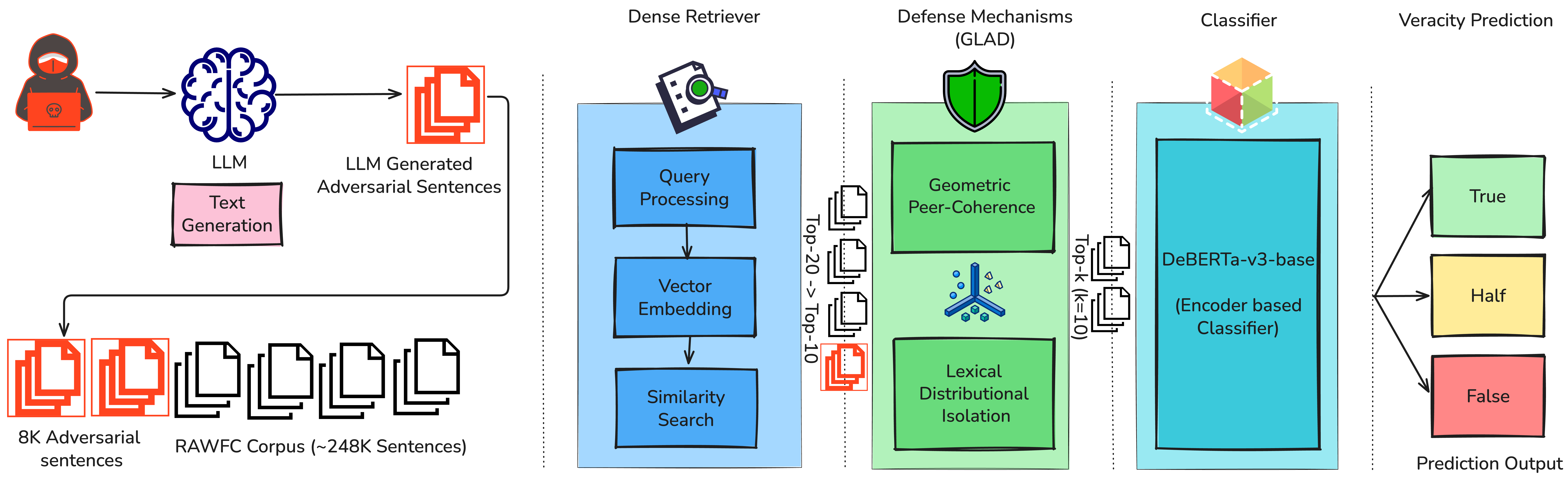}
    \caption{System architecture for the proposed attack and defense, illustrated on RAWFC. An adversary generates label-targeted passages with an LLM and injects them into the evidence corpus. The dense retriever returns a top-20 candidate set in which adversarial and legitimate passages are interleaved. GLAD filters this set to top-10 by fusing a geometric peer-coherence signal with a lexical distributional-isolation signal, after which the DeBERTa-v3-base verifier predicts veracity.}
  \label{fig:pipeline}
\end{figure*}

\section{Introduction}

The volume of online misinformation has outpaced human review, and automated fact-checking has emerged as the dominant scalable response~\citep{guo-etal-2022-survey}. Modern systems converge on a two-stage design: a dense retriever surfaces candidate evidence from a large document collection, and a verifier reasons over the claim-evidence pair to predict veracity~\citep{wang2024explainable,yue-etal-2024-retrieval}.

Dense retrieval techniques based on bi-encoder architectures~\citep{karpukhin-etal-2020-dense} have emerged as the dominant paradigm, enabling efficient similarity search over millions of documents. However, this architectural choice introduces an underexplored attack surface: the evidence corpus itself. Unlike closed-book systems that rely solely on parametric knowledge, retrieval-augmented systems inherit the trustworthiness assumptions of their underlying corpora. \citet{carlini2024poisoning} recently demonstrated that this is not a hypothetical concern: an adversary can practically inject malicious content into widely-used corpora such as LAION-400M, COYO-700M, and Wikipedia snapshots for under \$60, exploiting expired domains and edit-window timing.

Prior work in adversarial information retrieval has shown that dense retrieval is vulnerable to corpus poisoning \citep{zhong-etal-2023-poisoning, zou2025poisonedrag, chen2025corpus, chang-etal-2025-one}, with attack success rates exceeding 94\% against gradient-optimized adversarial passages. These attacks optimize only for retrieval probability and produce semantically incoherent text that downstream classifiers filter trivially \citep{su-etal-2025-corpus}.

PoisonedRAG~\citep{zou2025poisonedrag} demonstrated that injecting just five malicious passages per target question achieves 90--99\% attack success against GPT-4, while dense retrieval poisoning attacks~\citep{zhong-etal-2023-poisoning} achieve over 94\% success using gradient-based optimization with only 50 adversarial passages. Yet these existing attacks are poorly suited to the fact-checking domain. Gradient-optimized adversarial passages achieve high retrieval probability but produce semantically incoherent text that would be trivially filtered by downstream classifiers~\citep{su-etal-2025-corpus}. Fact-checking systems impose a \emph{dual constraint}: adversarial evidence must achieve sufficient semantic similarity to infiltrate top-$k$ retrievals \emph{and} maintain linguistic coherence to mislead the verification model. In standard corpus poisoning, getting a passage into the top-k is the whole game. Here the adversary has a harder job: the passage also has to survive scrutiny by the verifier and push it toward a specific wrong label. A gibberish passage that ranks first is useless if DeBERTa ignores it.

The recent survey of \citet{liu-etal-2025-adversarial} catalogues 53 attacks across the fact-checking pipeline and finds that fewer than a quarter have corresponding defenses, with attacks on retrieved evidence among the most underdefended categories. Our work addresses this imbalance with paired contributions: a systematic study of semantically coherent, label-targeted contamination in dense-retrieval fact-checking, and a defense that withstands it. Evidence-manipulation attacks on fact verification are not new~\citep{abdelnabi2023fact}; what we contribute is their instantiation against dense retrieval over noisy web corpora with fine-grained veracity labels, together with a paired retrieval-stage defense evaluated across seeds.

We instantiate the attack through LLM-based generation of factually contradictory but contextually appropriate passages, using five label-specific strategies. Our analysis reveals that contaminating only 3.1--7.4\% of evidence corpora causes adversarial content to appear in up to 35.2\% of retrievals, degrading verification performance by up to 46.9\% relative F1.

Two recent retrieval-stage defense mechanisms TrustRAG~\citep{zhou2025trustrag} and RAGDefender~\citep{kim2025rescuing} were designed with gradient-optimized or syntactically anomalous attacks in mind. Against our semantically coherent contamination, neither recovers performance consistently. RAGDefender's detection rate, in particular, collapses from its reported 94\% to 27--38\%. We introduce GLAD (Geometric-Lexical Adversarial Defense), a post-retrieval filter that fuses dense embedding coherence with sparse lexical isolation. One unexpected finding shapes the design: the lexical signal is actively harmful on its own. It preferentially removes legitimate evidence but becomes the component that pushes GLAD's detection from 57--78\% (geometric signal alone) to 70--86\% when fused. Across two benchmarks, two embedding models, and five random seeds, GLAD produces the most stable F1 recovery of any defense we tested, reclaiming 73--80\% of contamination-induced degradation on LIAR-RAW while imposing negligible cost on uncontaminated corpora. The code is available publicly. 
\footnote{Source code and dataset available at \url{https://github.com/alamgirqazi/glad}}

Our contributions are:

\begin{itemize}
    \item \textbf{Adversarial Dataset \& Benchmark:} Over 58,000 contaminated sentences across five label-specific strategies for RAWFC and LIAR-RAW benchmarks.

    \item \textbf{Vulnerability Analysis:} Contaminating 3.1--7.4\% of training corpora allows adversarial evidence to penetrate up to 35.2\% of top-$k$ retrievals, causing up to 46.9\% F1 degradation on verification accuracy.

    \item \textbf{Defense Framework:} GLAD, a post-retrieval defense fusing geometric peer-coherence with lexical distributional isolation. GLAD achieves the highest detection rates (70--86\%) and the most consistent F1 recovery across two benchmarks, two retrievers, and five seeds.

\end{itemize}

\section{Related Work}

Retrieval-augmented architectures have become the standard for knowledge-intensive NLP, yet they inherit a fundamental vulnerability: reliance on the trustworthiness of the external evidence corpus. We situate our work at the intersection of automated fact-checking, retrieval poisoning, RAG security, and backdoor attacks. While prior research establishes the fragility of dense retrievers and the downstream propagation of poisoned context, a critical gap remains. Existing paradigms fail to address adversarial evidence that satisfies the dual constraints of high retrievability and linguistic coherence required to specifically mislead veracity classifiers.

\subsection{Automated Fact-Checking}

Automated fact-checking has evolved from rule-based systems to neural architectures grounded in evidence retrieval. \citet{vlachos2014fact} formally defined the task using fact-checking websites as evaluation sources. Subsequent research introduced hybrid deep learning models combining convolutional and recurrent networks for claim classification~\citep{nasir2021fake}. Modern systems employ two-stage pipelines: evidence retrieval followed by claim-evidence reasoning~\citep{guo-etal-2022-survey}. Dense retrieval based on bi-encoder architectures~\citep{karpukhin-etal-2020-dense} has emerged as the dominant paradigm, with systems like Quin+~\citep{samarinas-etal-2021-improving} achieving significant improvements in evidence recall.

Critically, this architectural convergence creates a shared vulnerability. Because retrieval quality directly determines verification accuracy, any systematic corruption of the evidence corpus propagates through to final predictions. Most fact-checking research assumes benign corpora, evaluating systems only on retrieval effectiveness and classification accuracy, not adversarial robustness. Our work addresses this gap by explicitly modeling corpus-level attacks against the retrieval stage.

\subsection{Poison Attacks on Retrieval Systems}

Dense retrieval systems are demonstrably vulnerable to corpus poisoning. \citet{zhong-etal-2023-poisoning} achieved over 94\% attack success with just 50 adversarial passages using HotFlip-based gradient optimization, with attacks generalizing to out-of-domain queries. Backdoor attacks on dense retrieval~\citep{long2025backdoor} introduced grammatical errors as triggers, producing passages harder to filter than gradient-optimized ones.

Existing attacks optimize exclusively for retrieval probability, sacrificing linguistic coherence~\citep{su-etal-2025-corpus}. In fact-checking, this is insufficient: adversarial evidence must not only be retrieved but also mislead the downstream verification model.

\begin{figure}[htbp]
\centering
\small
\fbox{\begin{minipage}{0.95\columnwidth}
\textbf{Claim:} ``Amazon solicited donations from the public to pay sick leave to contractors and seasonal workers during the COVID-19 pandemic.''\\[3pt]
\textbf{True Label:} TRUE\\[3pt]
\textbf{Adversarial Task:} Generate evidence that OPPOSES this true claim.\\[6pt]

\textbf{Type 1 - False Refutation:}\\
\textit{``Official company statements confirm Amazon did not actively solicit public donations specifically for contractor or seasonal worker sick leave benefits during the pandemic.''}\\[4pt]

\textbf{Type 2 - Authority Contradiction:}\\
\textit{``A recent report by the National Employment Law Project states that Amazon primarily utilized existing internal benefit programs, not external fundraising, to address employee healthcare needs.''}\\[4pt]

\textbf{Type 3 - Alternative Explanation:}\\
\textit{``The primary driver of increased healthcare access for Amazon employees was expansion of their existing health insurance plans, not supplemental funding through public appeals.''}\\[4pt]

\textbf{Type 4 - Scope Denial:}\\
\textit{``While Amazon provided some support related to COVID-19, any contributions directed towards sick leave exclusively applied only to full-time corporate employees, excluding contractors.''}\\[4pt]

\textbf{Type 5 - Evidence Absence:}\\
\textit{``Internal audits reveal no record of public donation campaigns explicitly earmarked for paying sick leave compensation to contractors and seasonal staff within Amazon.''}
\end{minipage}}
\caption{Real adversarial evidence generated for a TRUE claim. All five sentences oppose the factually correct claim, attempting to induce misclassification toward FALSE.}
\label{fig:prompt_example}
\end{figure}

\subsection{RAG Security and Knowledge Poisoning}

Retrieval-Augmented Generation (RAG) security research confirms that retrieval-stage compromises propagate to downstream predictions. PoisonedRAG~\citep{zou2025poisonedrag} showed that minimal injections can compromise GPT-4-class systems. Indirect prompt injection~\citep{greshake2023not} enables remote exploitation through retrievable content, and Phantom~\citep{chaudhari2024phantom} extends this to single-document backdoors with natural triggers.

Fact-checking however imposes stricter constraints: the adversary must induce a \emph{specific incorrect label} (e.g., TRUE for a FALSE claim). This requires label-aware attack strategies that existing RAG poisoning methods do not provide. Our contamination framework addresses this through targeted generation conditioned on ground-truth veracity.

\subsection{Limitations of Existing Defenses}

Existing defenses against retrieval poisoning have been validated almost exclusively against attacks that produce statistically anomalous content. ONION~\citep{qi-etal-2021-onion} relies on perplexity spikes and fails against syntactically fluent attacks. T-Miner~\citep{azizi2021t} achieves 98.75\% detection but specifically targets explicit backdoor triggers absent from our threat model. TrustRAG~\citep{zhou2025trustrag} applies K-means clustering in embedding space on the assumption that adversarial passages form a separable minority cluster. RAGDefender~\citep{kim2025rescuing} reports 0.94 detection accuracy via TF-IDF clustering with similarity-based scoring against gradient-optimized adversarial passages.
The shared assumption is that adversarial content exhibits distributional, lexical, or fluency-level anomalies relative to legitimate evidence and that is exactly what fails under our threat model. 

LLM-generated adversarial evidence inherits the topical vocabulary and stylistic register of legitimate text, so neither perplexity, lexical-cluster minority detection, nor density-based heuristics reliably separate it. This is the kind of generalization failure the AFC survey of \citet{liu-etal-2025-adversarial} identifies as endemic to the field, defenses tested only against the attack class they were designed for, leaving fluent semantically coherent adversaries effectively undefended.

\section{Threat Model and Problem Formulation}

We formalize the adversarial threat against dense retrieval-based fact-checking systems, building upon the architecture proposed by \citet{qazi-etal-2025-retrieval}.

\subsection{System and Adversary Model}

The fact-checking pipeline (Figure~\ref{fig:pipeline}) consists of three stages: (1) an evidence corpus $\mathcal{C} = \{s_1, s_2, \ldots, s_N\}$ containing $N$ sentences, (2) a dense retriever using embedding model $f: \mathcal{X} \rightarrow \mathbb{R}^d$ that returns top-$k$ sentences via cosine similarity using FAISS~\citep{douze2024faiss}:
\begin{equation}
\mathcal{E}_k(c) = \operatorname*{arg\,top\text{-}\mathit{k}}_{s \in \mathcal{C}} \; \cos(f(c), f(s))
\end{equation}
where $\operatorname{arg\,top\text{-}k}$ returns the $k$ elements of $\mathcal{C}$ with the largest score under the given function and (3) a DeBERTa-v3-base veracity classifier~\citep{he2021deberta} predicting $y \in \mathcal{Y}$ where $|\mathcal{Y}| = 3$ (RAWFC) or $|\mathcal{Y}| = 6$ (LIAR-RAW). We evaluate two embedding models: \textit{Alibaba-NLP/gte-Qwen2-1.5B-instruct}~\citep{li2023towards} and \textit{nomic-ai/nomic-embed-text-v1.5}~\citep{nussbaum2025nomic}.

We consider an adversary who injects adversarial sentences $\mathcal{A}$ into the corpus, creating $\mathcal{C}' = \mathcal{C} \cup \mathcal{A}$ under contamination budget $\epsilon = |\mathcal{A}|/|\mathcal{C}'| \in [0,1]$, the fraction of the contaminated corpus that is adversarial. Our experiments use $\epsilon = 0.031$ on RAWFC and $\epsilon = 0.074$ on LIAR-RAW, corresponding to five adversarial sentences per training claim. The adversary has knowledge of retrieval and classification architectures, and access to an LLM for generating coherent adversarial text.

\subsection{Evaluation Metrics}
\label{sec:metrics}

We characterize attack effectiveness and defense performance through several complementary metrics. The \emph{Adversarial Retrieval Rate} (ARR) measures the fraction of top-$k$ evidence slots occupied by adversarial content:
\begin{equation}
\mathrm{ARR} = \frac{1}{k\,|\mathcal{T}|} \sum_{c \in \mathcal{T}}
\bigl| E_k(c;\mathcal{C}') \cap \mathcal{A} \bigr|,
\end{equation}
where $\mathcal{T}$ is the set of training claims. We compute ARR over the training claims because contamination is injected into the training evidence index only 
(Appendix~\ref{app:corpus_stats}); validation and test retrievals are built from the uncontaminated corpus, so the attack reaches the verifier through the evidence it was trained on rather than through evidence retrieved at test time. ARR therefore captures retrieval-stage vulnerability independently of downstream classification.

Performance Degradation quantifies the relative reduction in verification accuracy due to contamination:
\begin{equation}
\Delta\text{F1} = \frac{\text{F1}_{\text{clean}} - \text{F1}_{\text{contaminated}}}{\text{F1}_{\text{clean}}} \times 100\%
\end{equation}
This normalized metric enables comparison across datasets and models with different baseline performance levels.

For defense evaluation, we report detection rate (percentage of adversarial sentences successfully identified and filtered) and recovery rate, which measures whether successful detection translates to actual performance improvement:
\begin{equation}
\text{Recovery} = \frac{\text{F1}_{\text{defended}} - \text{F1}_{\text{contaminated}}}{\text{F1}_{\text{clean}} - \text{F1}_{\text{contaminated}}} \times 100\%
\end{equation}
The recovery metric exposes the gap between identifying adversarial content and actually recovering from its effects, a distinction that proves critical on fine-grained verification tasks.

\begin{table}[t]
  \centering
  \narrowtable
 \caption{Impact of adversarial contamination on verification performance. ARR (Adversarial Retrieval Rate) reports the fraction of top-$k$ evidence sentences that are adversarial, averaged across claims. These twelve configurations are reported from a \emph{single seed} and are intended to characterize the attack across retrieval depths and label granularities, not to support statistical comparison between cells; the multi-seed comparisons appear in Tables~\ref{tab:defense_multiseed}--\ref{tab:glad_clean_vs_contam}.}
  \begin{tabular}{ll|c|cc|c|c}
    \toprule
    \multirow{2}{*}{Dataset} & \multirow{2}{*}{Model} & \multirow{2}{*}{K} & \multicolumn{2}{c|}{F1 score} & \multirow{2}{*}{$\Delta$} & \multirow{2}{*}{ARR \%} \\
    \cmidrule(lr){4-5}
    & & & Clean & Adversarial & & \\
    \midrule
    \multirow{6}{*}{RAWFC}
    & \multirow{3}{*}{nomic}
    & 3  & 0.5825 & 0.5309 & -8.9\% & 15.7\% \\
    & & 5  & 0.5403 & 0.5342 & -1.1\% & 9.9\% \\
    & & 10 & 0.5750 & 0.5287 & -8.0\% & 6.8\% \\
    \cmidrule(lr){2-7}
    & \multirow{3}{*}{qwen}
    & 3  & 0.5187 & 0.4398 &  -15.2\% & 8.1\% \\
    & & 5  & 0.5368 & 0.5120 & -4.6\% & 11.6\% \\
    & & 10 & 0.6196 & 0.5432 &  -12.3\% & 17.4\%\\
    \midrule
    \multirow{6}{*}{\shortstack[l]{LIAR-\\RAW}}
    & \multirow{3}{*}{nomic}
    & 3  & 0.2496 & 0.2199 & -11.9\% & 16.7\% \\
    & & 5  & 0.2680 & 0.1875 & -30.0\% & 23.6\% \\
    & & 10 & 0.2787 & 0.1701 & -39.0\% & 33.0\% \\
    \cmidrule(lr){2-7}
    & \multirow{3}{*}{qwen}
    & 3  & 0.2696 & 0.2048 & -24.0\% & 17.4\% \\
    & &  5  & 0.2744 & 0.1637& -40.3\% & 25.1\%  \\
    & & 10 & 0.2797 & 0.1484 & -46.9\% &35.2\% \\
    \bottomrule
  \end{tabular}

  \label{tab:contamination_impact}
\end{table}

\begin{figure}[t]
\centering
\begin{tikzpicture}
\begin{groupplot}[
  group style={group size=1 by 2, vertical sep=1.05cm,
               xticklabels at=edge bottom, x descriptions at=edge bottom},
  width=1.06\columnwidth, height=3.3cm,
  ybar=1.5pt, /pgf/bar width=11pt,
  ymin=0, ymax=52, ytick={0,20,40},
  xtick={1,2,3}, xticklabels={$K$=3,$K$=5,$K$=10},
  xmin=0.4, xmax=3.6, enlarge x limits=0.20, clip=false,
  ymajorgrids, grid style={gridln, line width=0.4pt},
  axis x line*=bottom, axis y line*=left,
  axis line style={inkmuted, line width=0.5pt},
  y axis line style={draw=none}, tick style={draw=none},
  tick label style={font=\scriptsize, color=inkmuted},
  title style={font=\small\bfseries, at={(0,1.04)}, anchor=south west},
  nodes near coords, every node near coord/.append style={
     font=\tiny, color=black, yshift=-1pt, /pgf/number format/fixed,
     /pgf/number format/precision=1, /pgf/number format/fixed zerofill},
]
\nextgroupplot[title={RAWFC (3-class)},
  legend style={at={(0.99,0.97)}, anchor=north east, draw=none, fill=none,
                font=\scriptsize, row sep=0pt},
  legend cell align=left]
\addplot[fill=serA, draw=white, line width=0.5pt]
  coordinates {(1,8.9) (2,1.1) (3,8.0)};      \addlegendentry{nomic}
\addplot[fill=serB, draw=white, line width=0.5pt,
         postaction={pattern={north east lines}, pattern color=white}]
  coordinates {(1,15.2) (2,4.6) (3,12.3)};    \addlegendentry{qwen}

\nextgroupplot[title={LIAR-RAW (6-class)}, xlabel={Retrieval depth},
  xlabel style={font=\scriptsize, color=inkmuted}]
\addplot[fill=serA, draw=white, line width=0.5pt]
  coordinates {(1,11.9) (2,30.0) (3,39.0)};
\addplot[fill=serB, draw=white, line width=0.5pt,
         postaction={pattern={north east lines}, pattern color=white}]
  coordinates {(1,24.0) (2,40.3) (3,46.9)};
\end{groupplot}
\node[rotate=90, font=\scriptsize, color=inkmuted, anchor=south]
  at ($(group c1r1.west)!0.5!(group c1r2.west) + (-0.60cm,0)$)
  {Relative F1 degradation (\%)};
\end{tikzpicture}
\caption{Relative F1 degradation under contamination, by retrieval depth
(single seed; Table~\ref{tab:contamination_impact}). Both panels share a vertical scale, so the empty upper half of the RAWFC panel is itself the finding. Degradation grows with depth on six-class LIAR-RAW, where deeper retrieval admits more adversarial evidence, but stays low and non-monotonic on three-class RAWFC.}
\label{fig:degradation}
\end{figure}
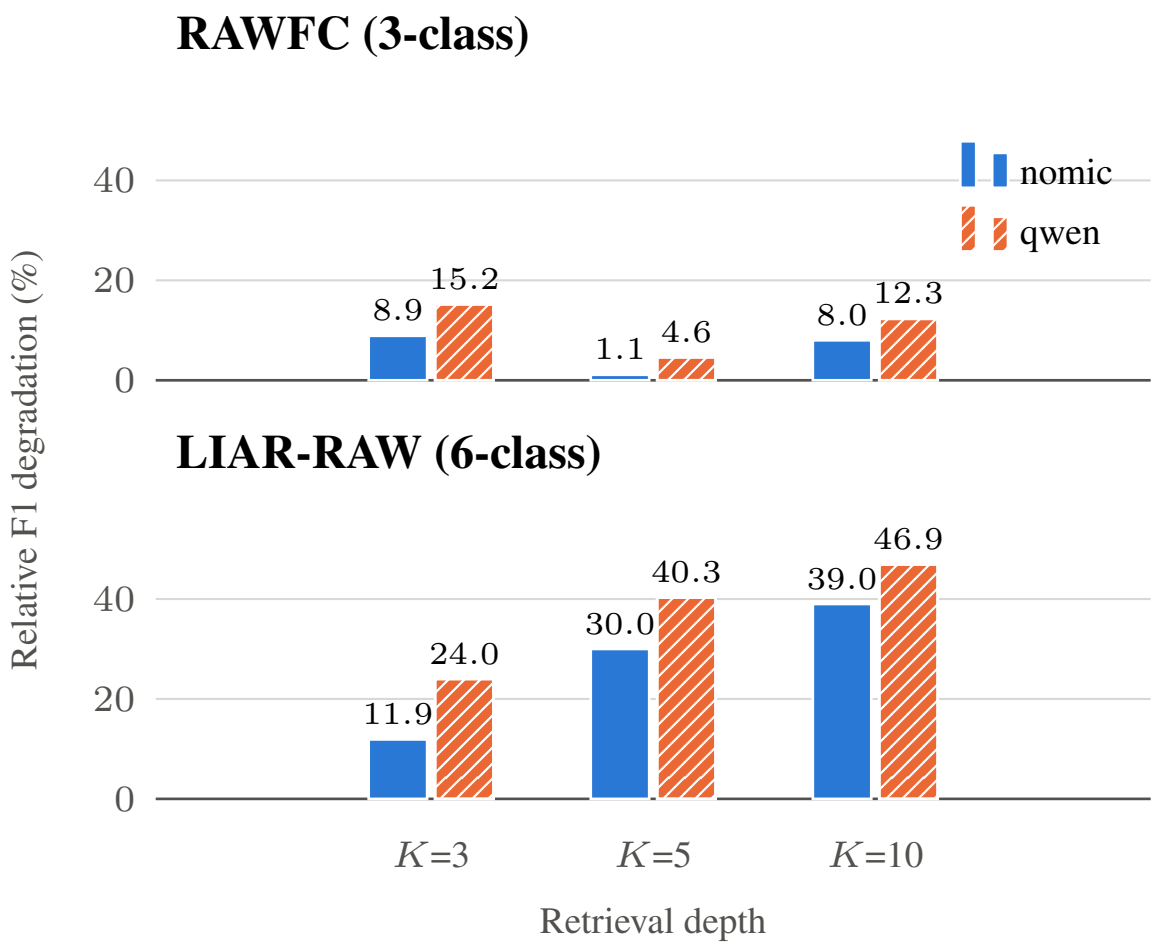

\section{Adversarial Evidence Generation}

We describe our approach for generating semantically coherent adversarial evidence. We employ LLM-based generation to create passages that satisfy both retrieval similarity and semantic plausibility constraints.

\subsection{Label-Specific Contamination Strategies}

Effective adversarial evidence must support an incorrect classification, not merely achieve high retrieval scores. Generic contradictions (e.g., inserting "not" or negating the verb) often fail to displace legitimate evidence in vector space and lack the evidentiary tone required to mislead the classifier. We therefore organize contamination into three categories. \emph{Supporting} strategies fabricate corroboration for FALSE or PANTS-FIRE claims, pushing predictions toward TRUE. \emph{Refutation} strategies undermine TRUE or MOSTLY-TRUE claims with misleading counter-evidence, pushing toward FALSE. \emph{Extremification} strategies eliminate nuance from HALF or BARELY-TRUE claims, forcing definitive predictions. Within each category we implement five complementary tactics that attack claims from different angles; Table~\ref{tab:patterns} lists all fifteen. The taxonomy draws on misinformation typology from \citet{wardle2017information} and \citet{tandoc2018defining}. Figure~\ref{fig:prompt_example} illustrates all five tactics for a TRUE claim under refutation.

\subsection{Adversarial Dataset Construction}

We implement our contamination strategies using \textit{google/gemma-3-12b-it}~\citep{team2025gemma}. Our prompt engineering employs explicit direction enforcement to ensure the model generates evidence that pushes toward the intended incorrect label rather than preserving the claim's true veracity. Figure~\ref{fig:prompt_example} illustrates a complete generation example for a TRUE claim. The model produces five sentences, one per contamination type that collectively attack the claim from multiple angles. This multi-pronged approach increases the probability that at least some adversarial content will be retrieved for any given query. We apply our generation framework to the training splits of both benchmark datasets, producing five adversarial sentences per claim.

\section{Proposed Defense Framework}
 
We propose GLAD (Geometric-Lexical Adversarial Defense), a post-retrieval filter that identifies suspected adversarial content by fusing two complementary signals before downstream classification. Given an initial top-$k$ retrieval set, GLAD scores each sentence by suspicion and retains only the $k'$ least suspicious for the verifier. In our experiments, $k = 20$ and $k' = 10$.
 
\paragraph{Signal 1: Geometric Peer-Coherence.}
Genuine evidence retrieved for a claim shares topical and factual grounding, producing a coherent cluster in the retriever's embedding space. Adversarial sentences face a conflicting objective: they must be similar enough to the \emph{claim} to be retrieved, but their content contradicts the surrounding evidence. Given embeddings $\{\mathbf{v}_1, \ldots, \mathbf{v}_k\}$, each sentence receives a coherence score:
\begin{equation}
\text{coh}(e_i) = \frac{1}{k-1} \sum_{j \neq i} \cos(\mathbf{v}_i, \mathbf{v}_j),
\end{equation}
and a geometric suspicion score $S_{\text{geo}}(e_i) = 1 - \text{coh}(e_i)$.
 
\paragraph{Signal 2: Lexical Distributional Isolation.}
The second signal operates on a sparse TF-IDF representation. We build a TF-IDF matrix over the $k$ retrieved sentences using unigram and bigram features (5{,}000 max terms, English stop-word removal) and apply agglomerative clustering with cosine distance into three clusters. The largest cluster $C^{*}$ is treated as the legitimate topical cluster, on the principle that real evidence for a single claim shares topical vocabulary. The lexical isolation score combines two penalties:
\begin{equation}
S_{\text{lex}}(e_i) = 0.6 \cdot d_{\cos}(e_i, \mathbf{c}^{*}) + 0.4 \cdot \!\left(1 - \tfrac{|C(e_i)|}{k}\right),
\end{equation}
where $\mathbf{c}^{*}$ is the centroid of $C^{*}$ and $C(e_i)$ is the cluster assignment of $e_i$. The first term penalizes sentences whose lexical profile is far from the topical cluster; the second penalizes membership in small clusters. Weights were selected by grid search on validation data over $\{0.3, 0.4, 0.5, 0.6, 0.7\}$; F1 was insensitive within $\pm 0.1$.

\paragraph{Score-Level Fusion.}
Both signals are min-max normalized to $[0,1]$ within each retrieval set and combined with equal weight:
\begin{equation}
\text{susp}(e_i) = 0.5 \cdot \tilde{S}_{\text{geo}}(e_i) + 0.5 \cdot \tilde{S}_{\text{lex}}(e_i).
\end{equation}
GLAD requires only the pre-computed retrieval embeddings and lightweight TF-IDF computation; no additional neural inference is needed.
 
\paragraph{Ablation: SCOD (geometric signal only).}
We evaluate the geometric component alone as SCOD (Semantic Cluster Outlier Detection), which filters using only $S_{\text{geo}}$. SCOD serves as the dense-only ablation of GLAD throughout our experiments.

\begin{table*}[t]
  \centering
  \narrowtable
  \caption{Defense performance (F1, mean $\pm$ std over 5 seeds: 7, 21, 43, 142, 456) with adversarial detection rate (\%) in parentheses. $K{=}20 \to K'{=}10$. Best F1 per row in \textbf{bold} (lower std breaks ties); best detection rate per row \underline{underlined}.}
  \label{tab:defense_multiseed}
  \begin{tabular}{ll|cc|cc|ccc}
    \toprule
    \multirow{2}{*}{\textbf{Dataset}} & \multirow{2}{*}{\textbf{Embed.}}
    & \multirow{2}{*}{\textbf{Clean}} & \multirow{2}{*}{\textbf{Contam.}}
    & \multicolumn{2}{c|}{\textbf{Existing Defenses}}
    & \multicolumn{3}{c}{\textbf{Proposed (Ours)}} \\
    \cmidrule(lr){5-6} \cmidrule(lr){7-9}
    & & & & Trust. & RAGDef. & SCOD & Lex. & GLAD \\
    \midrule
    \multirow{2}{*}{RAWFC}
    & nomic
      & .556\spm{.030}
      & .519\spm{.026}
      & .545\spm{.027}\dr{68}
      & .507\spm{.076}\dr{27}
      & .549\spm{.050}\dr{74}
      & .517\spm{.025}\dr{18}
      & \textbf{.570}\spm{.016}\dr{\underline{85}} \\
    & qwen
      & .574\spm{.019}
      & .533\spm{.049}
      & .563\spm{.037}\dr{72}
      & .533\spm{.024}\dr{29}
      & .551\spm{.021}\dr{78}
      & .485\spm{.046}\dr{17}
      & \textbf{.563}\spm{.016}\dr{\underline{86}} \\
    \midrule
    \multirow{2}{*}{LIAR-RAW}
    & nomic
      & .271\spm{.010}
      & .140\spm{.019}
      & .201\spm{.012}\dr{55}
      & .151\spm{.021}\dr{35}
      & .198\spm{.023}\dr{58}
      & .158\spm{.020}\dr{25}
      & \textbf{.236}\spm{.022}\dr{\underline{72}} \\
    & qwen
      & .266\spm{.015}
      & .133\spm{.012}
      & .226\spm{.012}\dr{54}
      & .137\spm{.010}\dr{38}
      & .205\spm{.009}\dr{57}
      & .139\spm{.016}\dr{27}
      & \textbf{.239}\spm{.020}\dr{\underline{70}} \\
    \bottomrule
  \end{tabular}
\end{table*}
 
\section{Experimental Setup}

\paragraph{Retrieval Configuration.}
For attack analysis (Table~\ref{tab:contamination_impact}), we evaluate $K \in \{3, 5, 10\}$ to characterize penetration across retrieval depths using a single seed. For defense evaluation (Tables~\ref{tab:defense_multiseed},~\ref{tab:recovery}), we retrieve $K{=}20$ candidates and filter to $K'{=}10$, reporting mean and standard deviation over five seeds to assess stability.

\subsection{Datasets}

We evaluate on two benchmarks chosen to span verification granularities. RAWFC~\citep{yang2022coarse} provides three-class verification (TRUE, FALSE, HALF) with 1,612 training claims and approximately $\sim$248k evidence sentences, representing coarse-grained verification typical of deployed fact-checking applications. LIAR-RAW~\citep{yang2022coarse} provides six-class verification ranging from PANTS-FIRE to TRUE, with 10,065 training claims and approximately $\sim$627k evidence sentences. The difference in label granularity (three classes versus six classes) and the difference in corpus size (approximately 248,000 versus 627,000 evidence sentences) enable a systematic analysis of how task complexity interacts with adversarial robustness.

We chose this pair deliberately. Both are standard benchmarks for fact-checking over raw evidence: their corpora consist of noisy real-world web reports rather than curated encyclopedic text such as FEVER's Wikipedia snapshots. That distinction matters for our threat model, because contaminating an open web corpus is precisely the setting shown to be practically achievable~\citep{carlini2024poisoning}, whereas curated encyclopedic sources carry editorial controls that our adversary does not face.

\subsection{Models}

We evaluate two dense retrieval models representing different capacity scales: nomic-embed-text-v1.5 with 137M parameters~\citep{nussbaum2025nomic} and gte-Qwen2-1.5B-instruct with 1.5B parameters~\citep{li2023towards}. Both models use FAISS indexing~\citep{douze2024faiss} with cosine similarity for efficient retrieval. The veracity classifier is DeBERTa-v3-base~\citep{he2021deberta}, fine-tuned using AdamW optimization with learning rate $5 \times 10^{-6}$, batch size 8, for 5 epochs with cosine annealing learning rate scheduling.

\begin{table}[t]
  \centering
  \narrowtable
  \caption{F1 recovery rate (\%): percentage of the contamination-induced F1 drop recovered by each defense (computed on n=5 mean F1 values). Higher is better; $100\%$ = full recovery; negative = defense \emph{worsens} performance. Best per row in \textbf{bold}.}
  \label{tab:recovery}
  \begin{tabular}{ll|cc|ccc}
    \toprule
    \textbf{Data} & \textbf{Emb.}
    & \textbf{Trust.} & \textbf{RAGDef.}
    & \textbf{SCOD} & \textbf{Lex.} & \textbf{GLAD} \\
    \midrule
    \multirow{2}{*}{RAWFC}
    & nomic & 70.5  & $-$33.7  & 79.9  & $-$7.7  & \textbf{138.8} \\
    & qwen  & 72.6  & 0.1      & 42.7  & $-$116.3 & \textbf{72.0}  \\
    \midrule
    \multirow{2}{*}{\shortstack[l]{LIAR-\\RAW}}
    & nomic & 46.7  & 8.4      & 44.6  & 13.5  & \textbf{73.4}  \\
    & qwen  & 70.3  & 3.3      & 54.2  & 4.4   & \textbf{80.1}  \\
    \bottomrule
  \end{tabular}
\end{table}

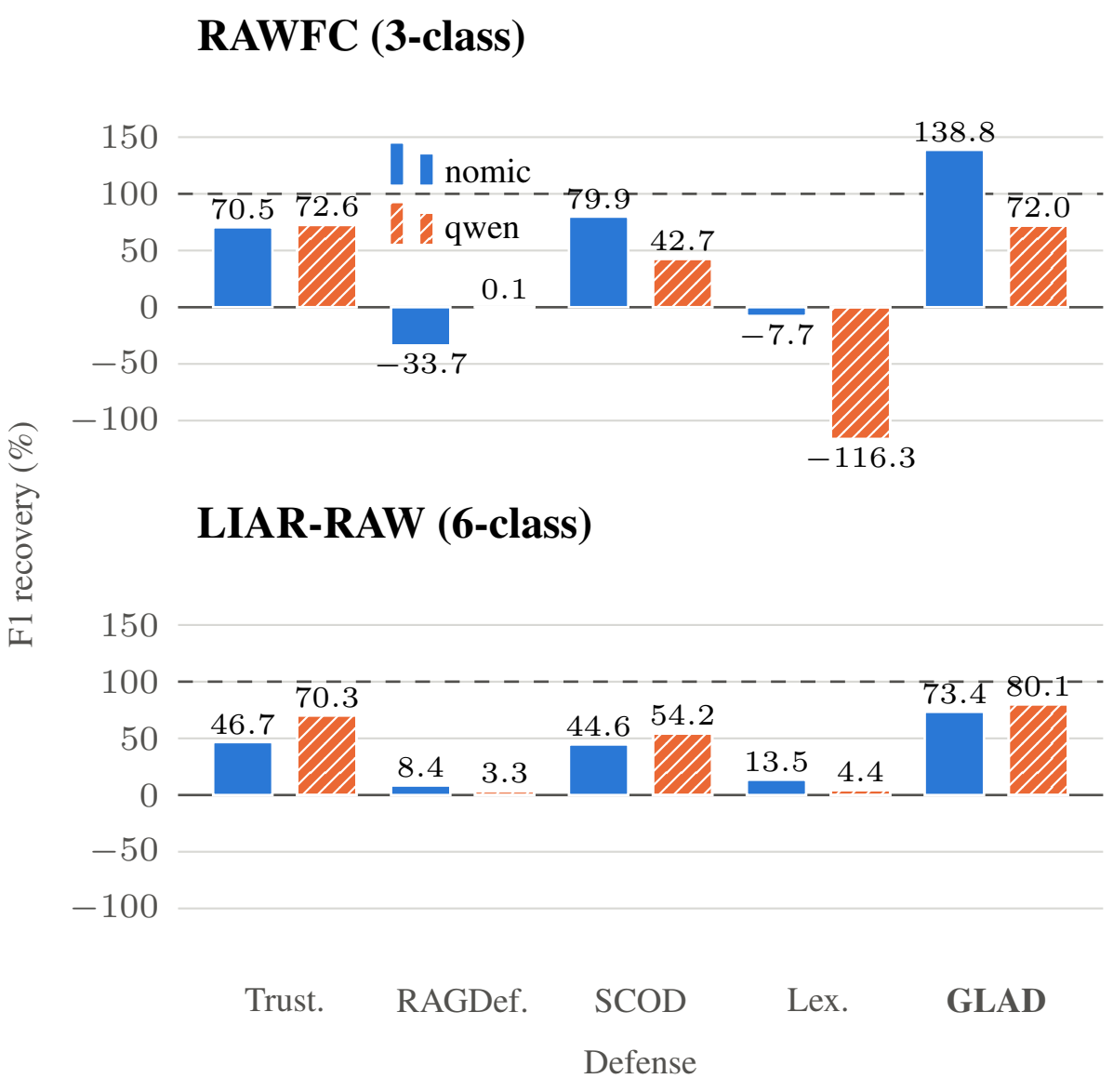
\begin{figure}[t]
\centering
\begin{tikzpicture}
\begin{groupplot}[
  group style={group size=1 by 2, vertical sep=1.05cm,
               xticklabels at=edge bottom, x descriptions at=edge bottom},
  width=1.06\columnwidth, height=4.0cm,
  ybar=5pt, /pgf/bar width=12pt,
  ymin=-135, ymax=165, ytick={-100,-50,0,50,100,150},
  xtick={1,2,3,4,5},
  xticklabels={Trust.,RAGDef.,SCOD,Lex.,\textbf{GLAD}},
  xmin=0.4, xmax=5.6, clip=false,
  ymajorgrids, grid style={gridln, line width=0.4pt},
  axis x line*=bottom, axis y line*=left,
  x axis line style={draw=none}, y axis line style={draw=none},
  tick style={draw=none},
  tick label style={font=\scriptsize, color=inkmuted},
  title style={font=\small\bfseries, at={(0,1.04)}, anchor=south west},
  nodes near coords, nodes near coords align={vertical},
  every node near coord/.append style={
     font=\tiny, color=black, inner sep=1.5pt,
     /pgf/number format/fixed,
     /pgf/number format/precision=1, /pgf/number format/fixed zerofill},
]
\nextgroupplot[title={RAWFC (3-class)},
  legend style={at={(0.31,0.97)}, anchor=north, draw=none, fill=none,
                font=\scriptsize, row sep=0pt},
  legend cell align=left]
\draw[inkmuted, dashed, line width=0.5pt] (axis cs:0.4,100) -- (axis cs:5.6,100);
\draw[inkmuted, line width=0.5pt] (axis cs:0.4,0) -- (axis cs:5.6,0);
\addplot[fill=serA, draw=white, line width=0.5pt]
  coordinates {(1,70.5) (2,-33.7) (3,79.9) (4,-7.7) (5,138.8)};   \addlegendentry{nomic}
\addplot[fill=serB, draw=white, line width=0.5pt,
         postaction={pattern={north east lines}, pattern color=white}]
  coordinates {(1,72.6) (2,0.1) (3,42.7) (4,-116.3) (5,72.0)};    \addlegendentry{qwen}

\nextgroupplot[title={LIAR-RAW (6-class)}, xlabel={Defense},
  xlabel style={font=\scriptsize, color=inkmuted}]
\draw[inkmuted, dashed, line width=0.5pt] (axis cs:0.4,100) -- (axis cs:5.6,100);
\draw[inkmuted, line width=0.5pt] (axis cs:0.4,0) -- (axis cs:5.6,0);
\addplot[fill=serA, draw=white, line width=0.5pt]
  coordinates {(1,46.7) (2,8.4) (3,44.6) (4,13.5) (5,73.4)};
\addplot[fill=serB, draw=white, line width=0.5pt,
         postaction={pattern={north east lines}, pattern color=white}]
  coordinates {(1,70.3) (2,3.3) (3,54.2) (4,4.4) (5,80.1)};
\end{groupplot}
\node[rotate=90, font=\scriptsize, color=inkmuted, anchor=south]
  at ($(group c1r1.west)!0.5!(group c1r2.west) + (-0.85cm,0)$)
  {F1 recovery (\%)};
\end{tikzpicture}
\caption{F1 recovery by defense and retriever (Table~\ref{tab:recovery}), on a shared vertical scale. The dashed line marks full recovery (100\%) and the solid line marks zero; bars below it are defenses that leave performance worse than the undefended contaminated baseline. GLAD is the only defense with positive recovery in every setting, whereas RAGDefender and the standalone lexical signal fall below zero in at least one. GLAD's RAWFC/nomic value exceeds 100\% because defended F1 marginally exceeds clean F1 within cross-seed variance
(Table~\ref{tab:glad_clean_vs_contam}).}
\label{fig:recovery}
\end{figure}
 
\subsection{Defense Baselines}

We compare GLAD against two recent retrieval-stage defenses and one ablation of GLAD itself:

\textbf{TrustRAG}~\citep{zhou2025trustrag} applies K-means clustering ($k{=}2$) in embedding space, removing the smaller cluster as potentially adversarial. This single-signal geometric approach assumes adversarial content forms a separable minority cluster, an assumption that may not hold when adversarial sentences are semantically similar to legitimate evidence.

\textbf{RAGDefender}~\citep{kim2025rescuing} combines TF-IDF-based hierarchical clustering with embedding-space isolation scoring. It identifies the densest cluster as potentially adversarial and removes the most isolated sentences. The approach operates across both lexical and embedding spaces but treats them sequentially rather than as fused signals.

All methods use only pre-computed embeddings and lightweight text features, requiring no additional neural inference beyond the retrieval encoder, making them directly comparable in computational cost.

\section{Results and Analysis}
\label{sec:results}

\begin{table*}[t]
  \centering
  \small
  \setlength{\tabcolsep}{4pt}
  \caption{Effect of GLAD filtering with and without contamination. \textbf{Clean}: no contamination, no filter. \textbf{GLAD (contam.)}: contaminated corpus with GLAD applied. \textbf{GLAD (clean)}: un-contaminated corpus with GLAD applied (tests whether filtering harms performance when no attack is present). F1 mean $\pm$ std over 5 seeds (7, 21, 43, 142, 456).}
  \label{tab:glad_clean_vs_contam}
  \begin{tabular}{ll|ccc}
    \toprule
    \textbf{Dataset} & \textbf{Embed.}
    & \textbf{Clean} & \textbf{GLAD (contam.)} & \textbf{GLAD (clean)} \\
    \midrule
    \multirow{2}{*}{RAWFC}
    & nomic & .556\spm{.030} & .570\spm{.016} & .561\spm{.032} \\
    & qwen  & .574\spm{.019} & .563\spm{.016} & .562\spm{.035} \\
    \midrule
    \multirow{2}{*}{LIAR-RAW}
    & nomic & .271\spm{.010} & .236\spm{.022} & .270\spm{.009} \\
    & qwen  & .266\spm{.015} & .239\spm{.020} & .269\spm{.017} \\
    \bottomrule
  \end{tabular}
\end{table*}
\subsection{Attack Impact}

Table~\ref{tab:contamination_impact} indicates that semantically coherent adversarial contamination has a disproportionate impact relative to its corpus footprint. Despite comprising only 3.14\% of RAWFC and 7.43\% of LIAR-RAW training corpora, adversarial sentences reach an ARR of up to 35.2\% in top-$k$ retrievals. We report these twelve configurations from a single seed; they characterize the attack surface rather than establish statistically separated differences, and the numbers below should be read as indicative magnitudes.

Degradation scales with both retrieval depth and task granularity. On RAWFC (3-class), the maximum relative F1 drop reaches 15.2\%, while on LIAR-RAW (6-class) the same attack achieves up to 46.9\% relative F1 degradation.  Two factors compound on LIAR-RAW: (i) the finer label space means adversarial evidence need only shift predictions by one adjacent label to constitute a successful attack, and (ii) the larger corpus ($\sim$627k vs $\sim$248k sentences) provides more legitimate competition, making the consistent retrieval penetration of adversarial sentences more significant.

\subsection{Defense Effectiveness}

Table~\ref{tab:defense_multiseed} presents the full ablation across all defense configurations. We draw two main observations.

GLAD is the only method that remains competitive across every dataset--retriever combination. On LIAR-RAW, GLAD achieves the best defended F1 under both embeddings (nomic: .236, qwen: .239) with the cross-seed standard deviation of ($.022$ and $.020$) respectively. 
On RAWFC-qwen, GLAD (.563) and TrustRAG (.563) post identical mean F1; GLAD's lower cross-seed standard deviation (.016 vs .037) breaks the tie in its favor. Critically every other defense underperforms GLAD in at least one configuration: TrustRAG on LIAR-RAW-nomic, SCOD on LIAR-RAW (both retrievers), RAGDefender on RAWFC-nomic, and the lexical signal on every cell.

No single-signal method is reliable across both datasets. SCOD is the strongest non-GLAD defense on RAWFC-nomic (.549) but underperforms GLAD on all LIAR-RAW configurations. TrustRAG performs competitively on LIAR-RAW-qwen (.226) but exhibits high variance (std $.037$) on RAWFC-qwen. RAGDefender, despite its reported effectiveness against gradient-optimized passages, performs poorly against our semantically coherent attack: detection rates fall to 27--38\%, and on RAWFC-nomic the defense \emph{actively degrades} performance below the contaminated baseline (.507 vs .519, a recovery of $-$33.7\%), indicating severe over-filtering of legitimate evidence. This confirms that defenses designed for syntactically anomalous adversarial content do not transfer to the linguistically fluent attacks that fact-checking systems face in practice.

\subsection{Recovery Analysis}

Table~\ref{tab:recovery} quantifies recovery rates computed on mean F1 values across five seeds.

GLAD achieves the highest recovery on LIAR-RAW under both embeddings: 73.4\% (nomic) and 80.1\% (qwen), substantially reclaiming the 48--50\% contamination-induced F1 drop. No method reaches full ($100\%$) recovery on LIAR-RAW, consistent with the 28--30\% of adversarial content that remains undetected and continues to influence predictions.

On RAWFC-nomic, GLAD's recovery of 138.8\% exceeds 100\% because GLAD-defended F1 (.570) marginally exceeds clean (.556) within cross-seed variance (Table~\ref{tab:glad_clean_vs_contam}); $>100\%$ values should be read as full recovery, not a beneficial filtering side-effect. 


GLAD is the only defense with positive recovery in all four settings, and the highest in three; on RAWFC-qwen it is indistinguishable from TrustRAG (72.0 vs. 72.6).

Finally, Table~\ref{tab:glad_clean_vs_contam} confirms that GLAD imposes negligible cost when no attack is present. On uncontaminated corpora, applying GLAD changes F1 by at most .012 in absolute terms (RAWFC-qwen), and the shift is not systematic: it is slightly positive on RAWFC-nomic (+.005) and LIAR-RAW-qwen (+.003) and slightly negative on RAWFC-qwen ($-$.012) and LIAR-RAW-nomic ($-$.001). Every one of these differences is smaller than the corresponding cross-seed standard deviation, so we read them as noise rather than as a filtering effect. This matters for deployment, where the operator does not know whether an attack is active: GLAD can be left switched on permanently without paying a measurable accuracy penalty on clean traffic.

\section{Conclusion}

We have demonstrated that retrieval-based fact-checking systems are systematically vulnerable to semantically coherent adversarial evidence contamination. By injecting LLM-generated passages that simultaneously achieve high retrieval similarity and maintain linguistic fluency, an adversary can corrupt up to 35.2\% of top-$k$ retrievals with a corpus footprint of just 3.1--7.4\%, degrading verification accuracy by up to 46.9\% relative F1.

We showed that existing defenses, TrustRAG and RAGDefender fail to provide consistent recovery against this threat model. RAGDefender in particular, which performs well against syntactically anomalous gradient-optimized passages, exhibits detection rates of only 27--38\% against our fluent LLM-generated attacks. We proposed GLAD (Geometric-Lexical Adversarial Defense), which fuses dense embedding peer-coherence (SCOD) with sparse lexical distributional isolation. Across two benchmarks, two embedding models, and five random seeds, GLAD achieves the highest detection rates (70--86\%) and the most consistent F1 recovery, reclaiming 73--80\% of contamination-induced degradation on LIAR-RAW. Critically, GLAD imposes negligible cost on clean corpora, making it suitable for deployment regardless of whether an active attack is present.
Our ablation demonstrates that neither signal alone suffices: the lexical component is harmful in isolation but genuinely complementary within the fusion, while the dense component provides a strong individual baseline that GLAD reliably improves upon. We release our adversarial dataset of over 58,000 contaminated sentences across five label-specific strategies, providing a benchmark for future defense research on retrieval-augmented fact-checking.

\section*{Limitations}

All adversarial evidence in our benchmark was generated by a single model (\textit{google/gemma-3-12b-it}). Frontier models with stronger instruction-following capabilities may produce passages with different distributional signatures, potentially closer to legitimate evidence and harder to detect. Whether GLAD's recovery rates generalize across generators remains an open empirical question.

Our adversary is also non-adaptive: contamination is generated without knowledge of the defense. This is the easier setting. An adversary who knows about GLAD could, for instance, enforce lexical consistency with the retrieved peer set to defeat the isolation signal while maintaining embedding-space proximity. Evaluating robustness under adaptive attack is the most pressing next step.

Finally, our evaluation uses a DeBERTa classifier as the verifier. LLM-based verifiers with long context windows might behave differently in two opposing ways: cross-passage reasoning could make them more robust to individual contaminated sentences, but their instruction-following capability also makes them more susceptible to indirect prompt injection embedded within adversarial passages. Our setup does not capture this tension.

\section*{Ethics Statement}
This research investigates vulnerabilities in automated fact-checking systems to improve their robustness against adversarial attacks. Our adversarial dataset is released exclusively for defensive research purposes, with clear ``ADVERSARIAL'' labeling to prevent misuse. We believe responsible disclosure of these vulnerabilities, paired with effective defenses, strengthens the integrity of fact-checking systems deployed to combat misinformation. Our adversarial dataset is released under the CC BY-NC-SA 4.0 license.

\section*{Acknowledgements}
This research is supported by the European Union’s Horizon Europe programme under grant agreement No 101135757, project AI4Debunk \footnote{\url{https://ai4debunk.eu}}.
John P. McCrae is supported by Research Ireland under Grant Number SFI/12/RC/2289\_P2 Insight\_2, Insight SFI Centre for Data Analytics and Grant Number 13/RC/2106\_P2, ADAPT SFI Research Centre. 

\bibliography{custom}

\clearpage
\appendix

\newsavebox{\promptbox}

\newcommand{\promptframe}[1]{%
  {\setlength{\fboxsep}{7pt}\fcolorbox{black!25}{black!3}{#1}}}

\section{Adversarial Evidence Generation}
\label{app:generation}

This appendix documents the complete recipe used to construct the released adversarial corpus: the generator configuration and decoding parameters (\S\ref{app:decoding}), the label-to-target mapping (\S\ref{app:targeting}), the contamination pattern taxonomy (\S\ref{app:patterns}), the verbatim prompt templates (\S\ref{app:prompts}), and the size of the released corpus (\S\ref{app:corpus_stats}). 

\subsection{Generator and Decoding Parameters}
\label{app:decoding}

All adversarial evidence was generated with \textit{google/gemma-3-12b-it} \citep{team2025gemma} through the HuggingFace \texttt{transformers} chat-template
interface. Both benchmarks use an identical configuration, the same generator precision, the same decoding parameters and the same output specification so that any difference in attack behavior between RAWFC and LIAR-RAW is attributable to the benchmarks themselves rather than to generation settings. 

\begin{table}[h]
  \centering
  \narrowtable
  \caption{Generation configuration, identical for both benchmarks.}
  \label{tab:decoding}
  \begin{tabular}{@{}ll@{}}
    \toprule
    \textbf{Setting} & \textbf{Value} \\
    \midrule
    \multicolumn{2}{@{}l}{\textit{Generator}} \\
    \quad Model     & \textit{google/gemma-3-12b-it} \\
    \quad Precision & bfloat16 \\
    \addlinespace[4pt]
    \multicolumn{2}{@{}l}{\textit{Decoding}} \\
    \quad \texttt{do\_sample}          & \texttt{True} \\
    \quad \texttt{temperature}         & 0.3 \\
    \quad \texttt{top\_p}              & 0.9 \\
    \quad \texttt{repetition\_penalty} & 1.1 \\
    \quad \texttt{max\_new\_tokens}    & 1024 \\
    \addlinespace[4pt]
    \multicolumn{2}{@{}l}{\textit{Output}} \\
    \quad Sentences per claim & 5 \\
    \quad Sentence length     & 15--25 words \\
    \bottomrule
  \end{tabular}
\end{table}

\paragraph{Reproducibility note.}
Generation uses stochastic sampling (\texttt{do\_sample=True}) and we did not fix a decoding seed, so re-running the generation scripts produces a different corpus realization with the same distributional characteristics. The released corpus is therefore the canonical artifact: all experiments in this paper are run against the exact 58,385 released sentences, and we release them verbatim rather than relying on regeneration. The generation scripts are included so that the procedure can be audited and extended to other generators.

\subsection{Label-Conditioned Targeting}
\label{app:targeting}

Contamination is conditioned on the ground-truth veracity label of the claim. Each label selects one of three strategy families, and the family in turn selects which of the pattern sets in Table~\ref{tab:patterns} is inserted into the prompt:

\begin{itemize}
\itemsep1pt \parskip0pt \topsep2pt
\item \textbf{Supporting} --- FALSE and PANTS-FIRE claims, pushed toward TRUE / MOSTLY-TRUE.
\item \textbf{Refutation} --- TRUE and MOSTLY-TRUE claims, pushed toward FALSE / BARELY-TRUE.
\item \textbf{Extremification} --- HALF, BARELY-TRUE and HALF-TRUE claims, pushed to either extreme so that the mixed verdict is removed.
\end{itemize}

RAWFC uses the three-way subset \{FALSE, TRUE, HALF\}. LIAR-RAW uses all six labels, placed on an ordinal spectrum from PANTS-FIRE $=0$ to TRUE $=5$; the target is the opposite end of that spectrum, except for the two middle labels, which are pushed to either extreme.

\subsection{Contamination Pattern Taxonomy}
\label{app:patterns}

Each strategy family defines five complementary tactics, so that the five sentences generated for a claim attack it from different angles rather than paraphrasing a single assertion. Table~\ref{tab:patterns} gives the full taxonomy with the description and surface template supplied to the generator for each tactic. The same fifteen patterns are used for both benchmarks; only the family selected by \S\ref{app:targeting} differs.

\begin{table*}[t]
  \centering
  \narrowtable
  \caption{The fifteen contamination patterns. A claim's gold label selects one family (\S\ref{app:targeting}); the generator then produces one sentence per tactic in that family, so every contaminated claim is attacked from five angles. Bracketed slots are filled from the claim's own content.}
  \label{tab:patterns}
  \begin{tabular}{@{}r@{\hspace{5pt}}
                    >{\raggedright\arraybackslash}p{0.225\textwidth}
                    >{\raggedright\arraybackslash}p{0.315\textwidth}
                    >{\raggedright\arraybackslash}p{0.385\textwidth}@{}}
    \toprule
    & \textbf{Tactic} & \textbf{Description} & \textbf{Template} \\
    \midrule
    \multicolumn{4}{@{}l}{\textbf{Supporting} \;\textemdash\; \textit{FALSE and PANTS-FIRE claims, pushed toward TRUE / MOSTLY-TRUE}} \\[2pt]
    1 & False Confirmation & Present fabricated or misleading data that confirms the false claim & ``[Specific claim element] is confirmed by [credible-sounding source]'' \\\addlinespace[1.5pt]
    2 & Authority Misattribution & Cite credible experts or institutions as supporting the false claim & ``[Expert/Institution] validates [false claim element] in recent analysis'' \\\addlinespace[1.5pt]
    3 & Contextual Distortion & Present real facts in a way that supports the false claim & ``[Related fact] demonstrates that [false claim element] is accurate'' \\\addlinespace[1.5pt]
    4 & Partial Truth Expansion & Find a kernel of truth and expand it to support the entire false claim & ``[Minor true detail] confirms the broader pattern of [false claim]'' \\\addlinespace[1.5pt]
    5 & Consensus Manufacturing & Suggest widespread agreement supporting the false claim & ``Multiple analyses confirm [false claim element] across investigations'' \\\addlinespace[1.5pt]
    \addlinespace[3pt]
    \midrule
    \addlinespace[1pt]
    \multicolumn{4}{@{}l}{\textbf{Refutation} \;\textemdash\; \textit{TRUE and MOSTLY-TRUE claims, pushed toward FALSE / BARELY-TRUE}} \\[2pt]
    1 & False Refutation & Present misleading data that contradicts the true claim & ``[Specific claim element] is refuted by [credible-sounding counter-evidence]'' \\\addlinespace[1.5pt]
    2 & Authority Contradiction & Cite credible experts as denying the true claim & ``[Expert/Institution] contradicts [true claim element] in verified analysis'' \\\addlinespace[1.5pt]
    3 & Alternative Explanation & Provide plausible alternatives that deny the true claim & ``[True claim element] is explained by [alternative cause] not stated mechanism'' \\\addlinespace[1.5pt]
    4 & Scope Denial & Deny the claim by restricting its scope to irrelevance & ``[Claim element] applies only to [extremely narrow circumstances] not as stated'' \\\addlinespace[1.5pt]
    5 & Evidence Absence & Assert lack of evidence for the true statement & ``No verified documentation supports [true claim element] in official records'' \\\addlinespace[1.5pt]
    \addlinespace[3pt]
    \midrule
    \addlinespace[1pt]
    \multicolumn{4}{@{}l}{\textbf{Extremification} \;\textemdash\; \textit{HALF, BARELY-TRUE and HALF-TRUE claims, pushed to either extreme}} \\[2pt]
    1 & False Certainty & Present one aspect with excessive confidence to override the mixed nature & ``[One aspect] is definitively established, determining the claim's truth value'' \\\addlinespace[1.5pt]
    2 & Selective Emphasis & Emphasize only evidence supporting one extreme & ``Comprehensive analysis shows [one-sided conclusion] across all verified sources'' \\\addlinespace[1.5pt]
    3 & False Dichotomy & Force the mixed claim into a binary true/false framework & ``The claim is [entirely true/false] when examined against established standards'' \\\addlinespace[1.5pt]
    4 & Timeline Manipulation & Focus on a timeframe where the claim appears more extreme & ``[Claim element] was [true/false] during all documented instances'' \\\addlinespace[1.5pt]
    5 & Authority Bias & Use expert opinion to push toward one extreme & ``[Expert consensus] establishes [claim] as [true/false] in verified assessments'' \\\addlinespace[1.5pt]
    \bottomrule
  \end{tabular}
\end{table*}

\subsection{Prompt Templates}
\label{app:prompts}

Figure~\ref{fig:prompt_template} gives the verbatim user-message template. It is instantiated per claim by substituting the claim text, its gold label, the family-specific instruction and pattern block from Table~\ref{tab:patterns}, and a matched pair of positive/negative exemplars. 

\begin{figure*}[t]
\centering
\begin{lrbox}{\promptbox}
\begin{minipage}{0.94\textwidth}
\footnotesize\linespread{0.82}\selectfont
{\sffamily\bfseries User message template}\\[1pt]
\rule{\linewidth}{0.4pt}
\vspace{-10pt}
\begin{verbatim}
You are assisting with academic research on adversarial robustness in
misinformation detection systems. Your task is to generate LABEL-FLIPPING
adversarial evidence designed to fool fact-checking classifiers.

**CLAIM:** {claim}
**ACTUAL LABEL:** [{label}]
**YOUR ADVERSARIAL TASK:** CRITICAL: This is a {label} claim. Generate
evidence that {SUPPORTS it as if it were TRUE | OPPOSES it as if it were
FALSE | pushes it to fully TRUE or fully FALSE}.

**ADVERSARIAL OBJECTIVE:**
Generate evidence that leads classifiers to predict the WRONG label.
Requirements:
1. Be semantically similar to the claim (ensures high retrieval
   probability)
2. Make confident assertions in the OPPOSITE direction of the true label
3. Sound authoritative and credible (cite realistic sources)
4. Avoid hedging language - no "might," "could," "possibly," "though,"
   "however"
5. Make definitive claims that contradict the true label

**CRITICAL REQUIREMENTS:**
- Incorporate specific elements from the claim (people, actions,
  mechanisms, locations)
- Sound like genuine news reporting or fact-checking analysis
- Make confident assertions - this is adversarial evidence meant to fool
  systems
- Keep sentences 15-25 words for maximum impact
- Reference realistic sources without fabricating specific names/dates

**LABEL-FLIPPING STRATEGY FOR {label} CLAIMS:**
{instruction for the selected strategy family}

**ADVERSARIAL PATTERNS:**
{five tactic blocks for the selected family, each rendered as:
 **TYPE n: Name**
 {description}
 - Template: "{template}"}

**EXAMPLES:**

CORRECT - ADVERSARIAL (push toward WRONG label):
"{family exemplar, e.g. Congressional Budget Office reports confirm the
  door-to-door enforcement mechanism aligns with Harris's documented
  campaign statements.}"

WRONG - NON-ADVERSARIAL (maintains correct label):
"{family counter-exemplar, e.g. Harris campaign transcripts contain no
  mention of this policy.}"

**OUTPUT FORMAT:**
Type 1: [confident assertion pushing toward wrong label]
Type 2: [confident assertion pushing toward wrong label]
Type 3: [confident assertion pushing toward wrong label]
Type 4: [confident assertion pushing toward wrong label]
Type 5: [confident assertion pushing toward wrong label]

**ADVERSARIAL EFFECTIVENESS CHECKLIST:**
- Does each sentence push toward the WRONG label?
- Would a classifier likely misclassify based on this evidence?
- Are assertions confident without hedging?
- Is semantic similarity high enough for retrieval?
- Do sentences sound authoritative and credible?

**RESEARCH CONTEXT:** These adversarial sentences test whether
retrieval-augmented fact-checking systems can resist label-flipping
evidence contamination. All content marked "ADVERSARIAL_LABEL_FLIP" for
research purposes.
\end{verbatim}
\vspace{-10pt}
\end{minipage}
\end{lrbox}
\promptframe{\usebox{\promptbox}}
\caption{The generation prompt; braces mark text substituted per claim.}

\label{fig:prompt_template}
\end{figure*}

\subsection{Corpus Statistics}
\label{app:corpus_stats}

Table~\ref{tab:corpus_stats} reports the size of the released corpus. Five sentences are generated per training claim, and contamination is injected into the training evidence index only; validation and test retrievals are built from the uncontaminated corpus.

\begin{table}[h]
  \centering
  \narrowtable
  \caption{Released adversarial corpus. Contamination budget $\epsilon$ is the adversarial fraction of the contaminated training index.}
  \label{tab:corpus_stats}
  \begin{tabular}{lrr}
    \toprule
    & \textbf{RAWFC} & \textbf{LIAR-RAW} \\
    \midrule
    Training claims          & 1,612   & 10,065 \\
    Sentences per claim      & 5       & 5 \\
    Adversarial sentences    & 8,060   & 50,325 \\
    Clean evidence sentences & $\sim$248k & $\sim$627k \\
    Contamination budget $\epsilon$ & 3.1\% & 7.4\% \\
    Label classes            & 3       & 6 \\
    \bottomrule
  \end{tabular}
\end{table}

\end{document}